\documentclass[aps,twocolumn,a4paper,superscriptaddress,nofootinbib]{revtex4-1}
\usepackage{graphicx,subcaption,amsmath,amsfonts,amssymb,multirow,extarrows,bm,acronym,float,caption}
\usepackage[colorlinks,linkcolor=blue,citecolor=blue,urlcolor=blue ]{hyperref}
\floatstyle{plaintop}
\restylefloat{table}
\newcommand{\PMO}{Key Laboratory of Dark Matter and Space Astronomy, Purple Mountain Observatory, Chinese Academy of Sciences, Nanjing 210023, People's Republic of China}
\newcommand{\USTC}{School of Astronomy and Space Science, University of Science and Technology of China, Hefei, Anhui 230026, People's Republic of China}

\def\apjl{Astrophysical Journal, Letters}             

\def\mnras{MNRAS}            
\def\prd{Phys.~Rev.~D}       
\def\prl{Phys.~Rev.~Lett}    

\begin{document}

\title{Tight constraints on Einstein-dilation-Gauss-Bonnet gravity from GW190412 and GW190814}

\author{Hai-Tian Wang}
\author{Shao-Peng Tang}
\affiliation{\PMO}
\affiliation{\USTC}
\author{Peng-Cheng Li}
\affiliation{Center for High Energy Physics, Peking University, No.5 Yiheyuan Rd, Beijing 100871, People's Republic of China}
\affiliation{Department of Physics and State Key Laboratory of Nuclear Physics and Technology, Peking University, No.5 Yiheyuan Rd, Beijing 100871, People's Republic of China}
\author{Ming-Zhe Han}
\author{Yi-Zhong Fan}
\email{Corresponding author: yzfan@pmo.ac.cn}
\affiliation{\PMO}
\affiliation{\USTC}

\date{\today}

\begin{abstract}
Gravitational-wave (GW) data can be used to test general relativity in the highly nonlinear and strong field regime.  
Modified gravity theories such as Einstein-dilation-Gauss-Bonnet and dynamical Chern-Simons can be tested with the additional GW signals detected in the first half of the third observing run of Advanced LIGO/Virgo. 
Specifically, we analyze gravitational-wave data of GW190412 and GW190814 to place constraints on the parameters of these two theories. 
Our results indicate that dynamical Chern-Simons gravity remains unconstrained. 
For Einstein-dilation-Gauss-Bonnet gravity, we find $\sqrt{\alpha_{\rm EdGB}}\lesssim 0.40\,\rm km$ when considering GW190814 data, assuming it is a black hole binary. 
Such a constraint is improved by a factor of approximately $10$ in comparison to that set by the first Gravitational-Wave Transient Catalog events. 
\end{abstract}

\maketitle

\acrodef{GW}{gravitational wave}
\acrodef{LIGO}{Laser Interferometer Gravitational-Wave Observatory}
\acrodef{LVC}{LIGO-Virgo Collaboration}
\acrodef{BNS}{binary neutron star}
\acrodef{NS}{neutron star}
\acrodef{BH}{black hole}
\acrodef{BBH}{binary black hole}
\acrodef{NSBH}{neutron star-black hole}
\acrodef{IMBH}{intermediate mass black hole}
\acrodef{GR}{general relativity}
\acrodef{PN}{post-Newtonian}
\acrodef{SNR}{signal-to-noise ratio}
\acrodef{PSD}{power spectral density}
\acrodef{PDF}{probability density function}
\acrodef{FIM}{fisher information matrix}
\acrodef{ppE}{parametrized post-Einsteinian}
\acrodef{IMR}{inspiral-merger-ringdown}
\acrodef{QNMs}{quasinormal modes}
\acrodef{ISCO}{innermost-stable circular orbit}
\acrodef{EdGB}{Einstein-dilation-Gauss-Bonnet}
\acrodef{dCS}{dynamical Chern-Simons}

\section{Introduction}\label{sec:intro}

Gravitational-waves (GWs) data have been widely used to test \ac{GR} in the strong gravity regime. 
No significant deviations from \ac{GR} have been found from analyses based on events from the first Gravitational-Wave Transient Catalog (GWTC-1) \citep{Test_GR_150914, TestGR_170817_PRL2019, Test_GR_PRD2019}.  
Recently, about $39$ GW events were reported by Advanced \ac{LIGO} and Advanced Virgo \citep{LIGO_O3a_arxiv2020}. 
Among them, there are three events/candidates involving the merger of one \ac{NS} or light mass \ac{BH} with another compact object (GW190425 \citep{GW190425_APJL_LIGO2020, GW190425_APJL_Han2020}, GW190426\_152155, and GW190814 \citep{GW190814_APJL_LIGO2020}), three GW events with mass ratios greater than $3$ (GW190412 \citep{GW190412_PRD_LIGO2020}, GW190426\_152155, and GW190814), and one merger producing an intermediate mass black hole (GW190521 \citep{GW190521_APJL_LIGO2020, GW190521_PRL_LIGO2020}). 
Based on these new data, dedicated analyses have been carried out to improve the general constraints on the deviations from \ac{GR} based on model independent tests \citep{Test_GR_arxiv2020}. 

Employing the publicly available posterior probability distributions released from the \ac{LVC} \citep{LIGO_PRX2019, LIGO_O3a_arxiv2020, Test_GR_PRD2019, Test_GR_arxiv2020}, one can set constraints on some specific modified gravity theories with the reweighting method. 
A more reliable method to constrain the theory is to perform a Bayesian analysis with modified GW waveform on real GW data.  
However, a full analytic \ac{IMR} waveform model is not available even in the \ac{GR} case since a prediction of the gravitational radiation during the merger stage is too complicated to be performed analytically. 
Currently, the merger stage is tractable only through numerical methods. 
The extension of such numerical methods to beyond \ac{GR} theories can be achieved only provided a well-behaved formulation of the equations of motion, which is presently lacking for many alternative theories in the strong-coupling limit \citep{Witek:2018dmd, Okounkova:2019dfo, Okounkova:2020rqw}. 
The inspiral stage is well described by \ac{PN} theory \citep{Blanchet_LRR2013} and the ringdown stage is fully described by a superposition of damped oscillations, i.e., the quasinormal modes \citep{QNM_LRR_Kokkotas1999, QNM_GRG_Ferrari2008, QNM_CQG_Berti2008}. 
For massive sources like GW190521, the signal detected by \ac{LIGO} and Virgo is dominated by the merger and ringdown stages. 
Instead, for most of other events, the inspiral stage is dominant since the masses of them are quite smaller than GW190521. 
By considering perturbative modifications to \ac{GR}, the \ac{ppE} framework \citep{Yunes_PRD2009, Yunes_PRD2016} has been developed to incorporate the effects of certain classes of modified gravity theories on GW signals (here, we focus on the inspiral stage). 

\ac{EdGB} \citep{EdGB_PRD_Kanti1996, EdGB_PRD_Yagi2016} and \ac{dCS} \citep{dCS_PRD_Alexander2009} are two particularly interesting classes of alternative theories to \ac{GR}.
Both of them have strong theoretical motivations (arising from, for instance, the low-energy limit of string theory or proposed to accommodate the matter-antimatter asymmetry of the Universe). 
The modifications to \ac{GR} are introduced by a scalar field nonminimally coupled to squared curvature scalars, which is why these theories are known as quadratic gravity theories. 
The nonvanishing scalar field results in a violation of the strong equivalence principle; thus, these two theories provide interesting models to test fundamental principles of \ac{GR}. 
By analyzing the events contained in the GWTC-1 catalog, \citet{EdGB_dCS_PRL_Nair2019} reported a constraint on \ac{EdGB}, $\sqrt{\alpha_{\rm EdGB}}\lesssim 5.6\,\rm km$, the first constraint on this parameter with GW data. 
A reliable constraint on \ac{dCS}, however, is not achievable with these events, yet. 

For GW events with a fixed \ac{SNR}, more stringent constraints on these two theories can be imposed if at least one of the components has a low mass. 
When only the secondary component mass of the binary is small, at current \ac{SNR}s, higher multipoles can have a sizable impact on the parameter estimation roughly for $q>3$ \citep{Pv3HM_PRD_Khan2020}. 
The analyses in \citet{GW190412_PRD_LIGO2020, GW190814_APJL_LIGO2020} also show that both the higher multipoles and precession will affect the estimation of masses and spins of the sources, which finally affect the constraints on \ac{EdGB} and \ac{dCS} gravity. 
Furthermore, there are subtleties in casting constraints on these theories by analysing \ac{NSBH} system. 
In \ac{EdGB} gravity, BHs have nonvanishing scalar charges, while NSs do not if the scalar field coupling is linear \citep{Modified_gravity_PRD_Yagi2012}, and the constraint on $\sqrt{\alpha_{\rm EdGB}}$ could be more promising for a mixed binary \citep{Mixed_binary_CQG_Carson2019}. 
For \ac{dCS} gravity, the leading order of the scalar charge effect enters at $2$ \ac{PN} while the tidal effect enters at $1$ \ac{PN} \citep{tidal_PRD_Vines2011, tidal_PED_Damour2012}, and the perturbations taking into account tidal effects have not yet been computed. 
So we do not perform analyses on \ac{dCS} theory for \ac{NSBH} cases. 

In this work, we first adopt the reweighting method to analyze GW data of two selected events of GWTC-2 
(i.e., GW190412 and GW190814), whose mass ratios are larger than $3$ and for which the inspiral \ac{SNR}s are larger than $6$ \citep{Test_GR_arxiv2020}. 
Then we use Bayesian inference method to analyze these two events. 
We focus on the modification due to \ac{EdGB} and \ac{dCS} in the inspiral phase of GW signal, within the \ac{ppE} framework. 
The methods we adopted are described in Sec.~\ref{sec:method}, and the main results are presented in Sec.~\ref{sec:result}. 
Finally, we summarize the results obtained in this work in Sec.~\ref{sec:summary}. 
We assume $G = c = 1$ throughout the paper, unless otherwise specified. 

\section{Methods}\label{sec:method}
For both \ac{EdGB} and \ac{dCS} theories, the modifications on the phase of the inspiral stage can be incorporated into the \ac{ppE} formalism, which describes the gravitational waveform of theories alternative to \ac{GR} in a model-independent way \citep{Yunes_PRD2009}. 
Indeed, the effects of the scalar dipole (in \ac{EdGB}) and scalar quadrupole (in \ac{dCS}) radiation are captured by the \ac{ppE} parameter entering at $-1$ and $2$ \ac{PN} order, respectively. 
Although these modifications affect the amplitudes of GW signals, the results presented in \citet{Tahura_PRD2019} show that phase-only analyses can place tight constraints. 
For a binary compact object with component masses $m_1$ and $m_2$ in the detector frame, the ending frequency of the inspiral stage can be approximated by $f_{\rm high}=0.018/[(m_1+m_2)]$ \citep{Test_GR_PRD2019}. 
Above this frequency the waveform model cannot be well described by the \ac{PN} expansion and the \ac{ppE} formalism and needs to be included in a different way \citep{Yunes_PRD2016}. 
Therefore, in this work, we only consider phase corrections and analyze GW data with inspiral waveform only, since the signal sourced by these events is dominated by the inspiral radiation. 

The inspiral stage of the IMR waveform model can be expressed in the form of $h_{\mathrm{ins}}(f)=A_{\mathrm{ins}}(f)e^{i{\Psi}_{\mathrm{ins}}}$; then, the waveform model including the modification due to \ac{EdGB} or \ac{dCS} can be written as $h_{\mathrm{ins}}(f)=A_{\mathrm{ins}}(f)e^{i{\Psi}_{\mathrm{ins}}+i\beta_{\rm EdGB,dCS}(\pi\mathcal{M}f)^{b}}$, where $b=-7/3$ ($b=-1/3$) for \ac{EdGB} (\ac{dCS}) gravity. 
The coupling constants in these two theories are $\alpha_{\rm EdGB}$ and $\alpha_{\rm dCS}$, respectively. 
For \ac{EdGB} gravity \citep{Modified_gravity_PRD_Yagi2012}, $\beta_{\mathrm{EdGB}}$ can be expressed as 
\begin{equation}
\label{eq:beta_EdGB}
\beta_{\mathrm{EdGB}}=-\frac{5}{7168} \frac{\left(m_{1}^{2} s_{2}^{\mathrm{EdGB}}-m_{2}^{2} s_{1}^{\mathrm{EdGB}}\right)^{2}}{\eta^{18 / 5}M^{4}} \frac{16 \pi \alpha_{\mathrm{EdGB}}^{2}}{M^{4}}.
\end{equation}
Here, $\eta=m_1m_2/M^2$ is the symmetric mass ratio, $M=m_1+m_2$ is the total mass, $\mathcal{M}=(m_1m_2)^{3/5}/(m_1+m_2)^{1/5}$ is the chirp mass, and $s_{i}^{\mathrm{EdGB}}=2\left(\sqrt{1-\chi_{i}^{2}}-1+\chi_{i}^{2}\right)/\chi_{i}^{2}$ are the dimensionless \ac{BH} charges in \ac{EdGB} gravity \citep{TestGR_GRG_Berti2018, EdGB_PRD_Prabhu2018}, where $\chi_i=\vec{S}_{i} \cdot \hat{L} / m_{i}^{2}$ are the dimensionless spins of \ac{BH}s with spin angular momentum $\vec{S}_{i}$ in the direction of the orbital angular momentum $\hat{L}$. 

For \ac{dCS} gravity, the phase modification takes the form \citep{dCS_PRL_Yagi2012} 
\begin{equation}
\label{eq:beta_dCS}
\begin{aligned}
\beta_{\mathrm{dCS}}=&\frac{16 \pi \alpha_{\mathrm{dCS}}^{2}}{M^{4} \eta^{14 / 5}}\biggl(-\frac{5}{8192} \frac{\left(m_{1} s_{2}^{\mathrm{dCS}}-m_{2} s_{1}^{\mathrm{dCS}}\right)^{2}}{M^{2}} \\
& +\left.\frac{15075}{114688}\left(\frac{m_{1}^{2} \chi_{2}^{2}+m_{2}^{2} \chi_{1}^{2}}{M^{2}}-\frac{350}{201} \eta \chi_{1} \chi_{2}\right)\right), 
\end{aligned}
\end{equation}
where $s_{i}^{\mathrm{dCS}}=\left[2+2\chi^4_i-2\sqrt{1-\chi_{i}^{2}}-\chi_{i}^{2}\left(3-2\sqrt{1-\chi^2_i}\right)\right]$
$/2\chi_{i}^{3}$ are the dimensionless \ac{BH} scalar charges of \ac{dCS} gravity \citep{Yunes_PRD2016}. 
Note that $s_{i}^{\mathrm{EdGB,\,dCS}}=0$ for ordinary stars like \ac{NS} \citep{Modified_gravity_PRD_Yagi2012, ppe_PRD_Tahura2018}. 
Therefore, to have a nonvanishing phase correction, at least one of the two compact objects involved in the merger should be a \ac{BH}. 
Among GWTC-2 events, GW190814 is well known for the large mass ratio, and the nature of the light object remains unclear. 
Though the NS nature cannot be ruled out, it is hard to understand why the first \ac{NSBH} merger event hosts a neutron star as massive as  about $2.6M_\odot$, which should be very rare even if such a group of heavy NSs does exist \citep{2021ApJ...908L...1T, 2021ApJ...908L..28N, 2020PhRvD.102f3006S}. 
Therefore, in this work, we assume that GW190814 is a \ac{BBH} event. 

A meaningful constraint on $\sqrt{\alpha_{\rm EdGB,dCS}}$ requires the dimensionless coupling constants $\zeta_{\rm EdGB,dCS}=16 \pi \alpha_{\mathrm{EdGB,dCS}}^{2}/m_s^4<1$, where $m_s$ is a typical mass scale of a system; here, we choose it to be the smaller mass scale of the components \citep{EdGB_dCS_PRL_Nair2019}. 
Then, we have\footnote{We do not set this constraint as a prior during the Bayesian inference below, but the posterior shows that the gathered samples do satisfy this condition.} 
\begin{equation}
\label{eq:threshold}
\sqrt{\alpha_{\rm EdGB,dCS}}\lesssim 0.74\frac{m_s}{M_{\odot}},
\end{equation}
where $M_{\odot}$ is the solar mass. 
So, events with lower component masses should place better constraints on both modified gravity theories.

We employ two methods to place constraints on these two modified gravity theories. 
As for the first one, we directly use the posteriors of $(\delta\phi_{4,-2}, m_1, m_2, a_1, a_2, \theta_1, \theta_2)$ released by the \ac{LVC}, where $\delta\phi_{4,-2}$ are obtained with the IMRPhenomPv3HM waveform model \citep{Pv3HM_PRD_Khan2020} in the work by \citet{Test_GR_arxiv2020} and the rest parameters are obtained with the IMRPhenomPv3HM waveform model in the work by \citet{LIGO_O3a_arxiv2020}. 
Among the above parameters, $\chi_i=a_i\cos\theta_i,\,i=1,2$, where $a_i$ are the dimensionless spin magnitudes, $\theta_{1,2}$ are angles between the component spins and the orbital angular momentum \citep{2015PhRvD..91d2003V}, and $\delta\phi_{4,-2}$ are expressed as
\begin{equation}
\label{eq:phi-24}
\begin{aligned}
\delta\phi_{-2}&=\frac{128\beta_{\rm EdGB}}{3\eta^{2/5}}, \\
\delta\phi_4&=\frac{128\eta^{4/5}\beta_{\rm dCS}}{3\phi_4},
\end{aligned}
\end{equation}
where $\phi_4$ is the \ac{GR} coefficient at $2$\ac{PN} order \citep{fd_PRD_Khan2016}. 
Combining Eq.(\ref{eq:phi-24}) with Eqs.(\ref{eq:beta_EdGB}) and (\ref{eq:beta_dCS}), we can get the posteriors of $\sqrt{\alpha_{\rm EdGB, dCS}}$ by reweighting, assuming a uniform prior on $\sqrt{\alpha_{\rm EdGB, dCS}}$. 

For the second one, we perform a Bayesian analysis to infer the posterior probability distributions of the source parameters $\mathcal{\theta}$, and the likelihood in this analysis is written as 
\begin{equation}
\label{eq:likelihood}
\begin{aligned}
p(d|\mathcal{\theta},S)\varpropto \mathrm{exp}\left[-\frac{1}{2}\sum^{N}_{i=1}\langle d_i-h_i|d_i-h_i\rangle\right],
\end{aligned}
\end{equation}
where $d_i$ is the data of the $i$th instrument, $S$ is the model assumption, $N$ is the number of detectors in the Advanced \ac{LIGO}/Virgo network, and $h_i$ is the waveform model projected on the $i$th detector. 
The noise weighted inner product $\langle a(f)|b(f)\rangle$ is defined by 
\begin{equation}
\label{eq:inner_product}
\langle a(f)|b(f)\rangle = 4\mathcal{R}\int^{f_{\rm high}}_{f_{\rm low}}\frac{a(f)b(f)}{S_n(f)} \mathrm{d}f, 
\end{equation}
where $f_{\rm low}$ and $f_{\rm high}=0.018/[(m_1+m_2)]$ are the high-pass and low-pass cutoff frequencies, respectively, and $S_n(f)$ is the power spectral density of the detector noise. 
The Bayes theorem is 
\begin{equation}
\label{eq:bayes}
p(\mathcal{\theta}|d,S) = \frac{p(d|\mathcal{\theta},S)p(\mathcal{\theta}|S)}{p(d|S)},
\end{equation}
where $p(d|S)=\int p(d|\mathcal{\theta},S)p(\mathcal{\theta}|S)\mathrm{d}\mathcal{\theta}$ is the evidence of the specific model assumed. 
The GW data and power spectral density for each event are downloaded from \ac{LIGO}-Virgo GW Open Science Center \citep{LIGO_O3a_arxiv2020}. 

To estimate the parameters with \ac{GW} data of GW190412 and GW190814, we carry out Bayesian inference with the {\sc Bilby} package \citep{Ashton_APJ2019} and {\sc Dynesty} sampler \citep{Dynesty_MNRAS_Speagle2020} with $1000$ live points.
The minimal (walks) and maximal (maxmcmc) number of sampling steps are $100$ and $5000$, respectively. 
We carry out Bayesian inference with the IMRPhenomXPHM waveform model \citep{XPHM_arxiv_Pratten2020} for both GW190412 and GW190814. 
Note that the IMRPhenomXPHM waveform model is more accurate and more efficient than the IMRPhenomPv3HM waveform model, which is used in reweighting analyses above. 
Thus, as a comparison, we also carry out Bayesian inference with the IMRPhenomPv3HM waveform model for GW190814. 
Because of a much smaller mass ratio of GW190412, we do not expect an obvious difference between results of the IMRPhenomPv3HM and IMRPhenomXPHM waveform models. 
So, the comparison between these two waveform models is absent for GW190412. 
The priors on $\sqrt{\alpha_{\rm EdGB}}$ and $\sqrt{\alpha_{\rm dCS}}$ are chosen to be uniform in the ranges of $[0,\,5]$ and $[0,\,50]\,\rm km$, respectively. 
For other parameters, we use the same priors as those used in Ref.~\citep{LIGO_O3a_arxiv2020}.
The high-pass (low-pass) cutoff frequencies $f_{\rm low\,(high)}$ of GW190412 and GW190814 are $20$ and $20\,{\rm Hz}$ ($83$ and $137\,{\rm Hz}$), respectively. 

\section{Results}\label{sec:result}
\begin{figure*}
\centering
\begin{subfigure}[b]{0.5\linewidth}
\centering
\includegraphics[width=\textwidth,height=8cm]{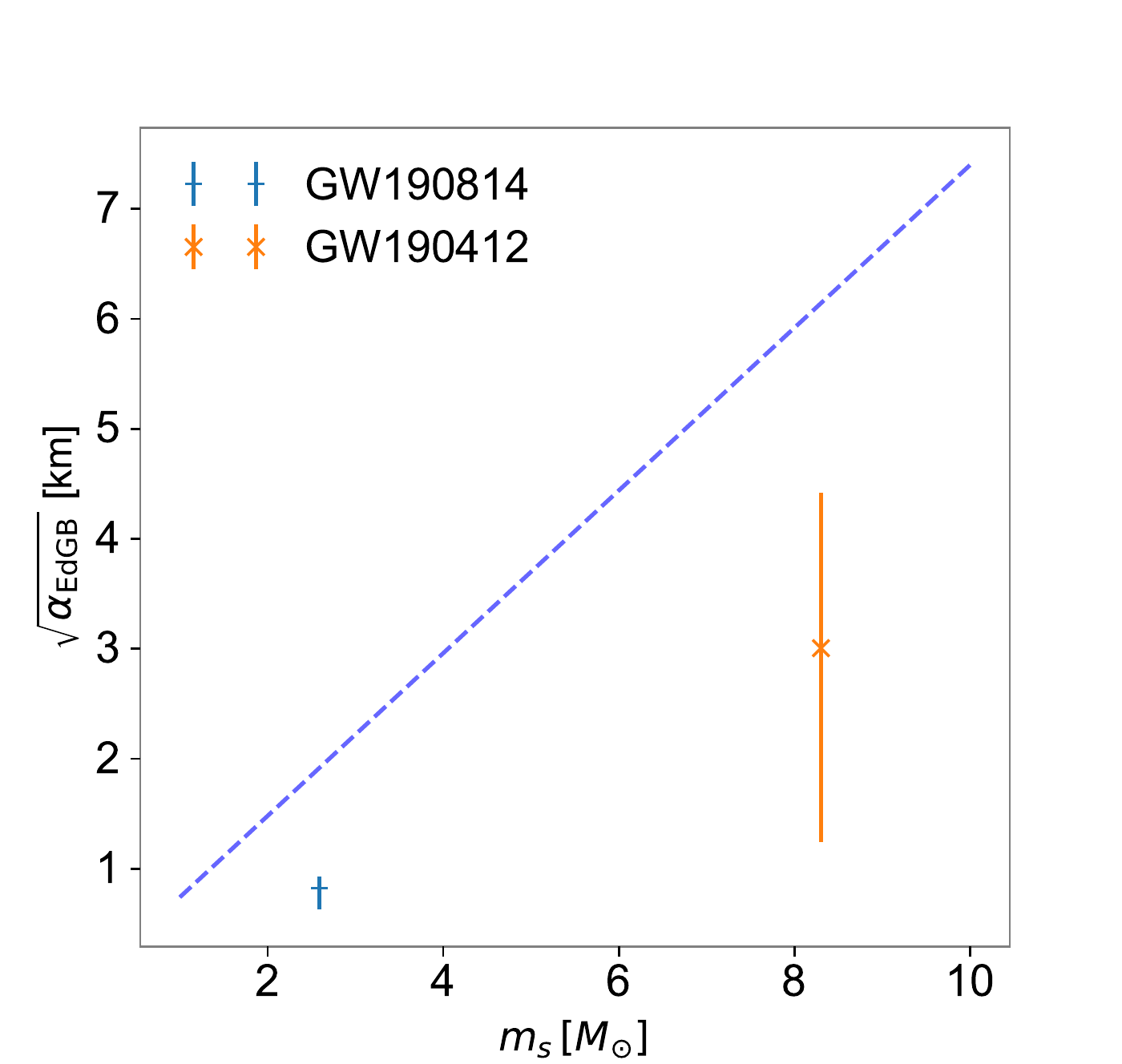}
\end{subfigure}%
\begin{subfigure}[b]{0.5\linewidth}
\centering
\includegraphics[width=\textwidth,height=8cm]{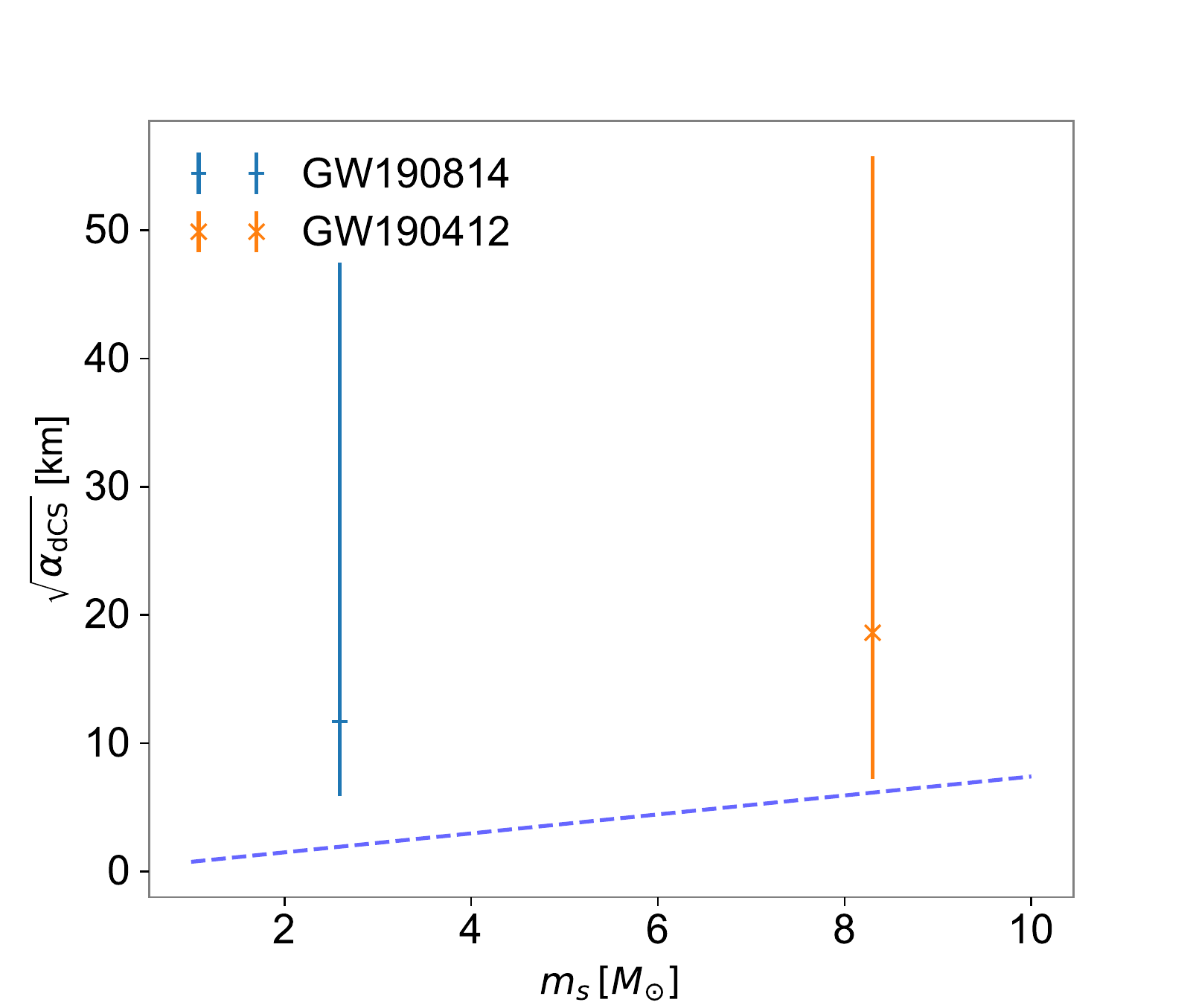}
\end{subfigure}%
\caption{
Reweighted results of $\sqrt{\alpha_{\rm EdGB}}$ (left panel) and $\sqrt{\alpha_{\rm dCS}}$ (right panel) for GW190412 and GW190814. 
The dashed lines are the relation between the median value of the secondary mass and the threshold value of $\sqrt{\alpha_{\rm EdGB,dCS}}$ given by Eq.(\ref{eq:threshold}). 
The secondary masses of GW190412 and GW190814 are $8.3^{+1.6}_{-0.9}M_{\odot}$ and $2.59^{+0.08}_{-0.09}M_{\odot}$ at $90\%$ credibility \citep{LIGO_O3a_arxiv2020}. 
Then the threshold values of $\sqrt{\alpha_{\rm EdGB,dCS}}$ are calculated with the median values of the secondary masses of these two events. 
The data points show the median values and $90\%$ regions of the $\sqrt{\alpha_{\rm EdGB,dCS}}$ posterior distributions. 
Evidently, both GW190814 and GW190412 can place meaningful constraints on $\sqrt{\alpha_{\rm EdGB}}$, and GW190814 works better than GW190412. 
However, both events are unable to place meaningful constraints on $\sqrt{\alpha_{\rm dCS}}$. 
}
\label{fig:EdGB_dCS_reweight}
\end{figure*}

By performing reweighting analysis with GW190412 and GW190814, we find that GW190814 yields the tightest constraint $\sqrt{\alpha_{\rm EdGB}}\lesssim 0.93 \,\rm km$ at $90\%$ credibility. 
The constraint from GW190412 is $\sqrt{\alpha_{\rm EdGB}}\lesssim 4.41\,\rm km$ at $90\%$ confidence level. 
Both events, however, cannot place useful constraint on $\sqrt{\alpha_{\rm dCS}}$ since most of the posterior distributions are significantly above the threshold value, as shown in Fig.\ref{fig:EdGB_dCS_reweight}.
With these results, it is reasonable to expect that GW190814 will impose a tighter constraints when employing a full Bayesian inference method. 

Recently, based on the general consideration about the stability of BHs in \ac{EdGB} gravity \citep{Witek:2018dmd}, \citet{QNM_arxiv_Carullo2021} has obtained the most stringent constraint  to date on \ac{EdGB}, $\sqrt{\alpha_{\rm EdGB}}\lesssim 0.8\,\rm km$, by assuming the secondary object of GW190814 is a \ac{BH}. 
\footnote{Note that Ref.~\citep{Witek:2018dmd} uses a different units convenient in the action of the \ac{EdGB} theory. 
To keep the convention of Ref.~\citep{EdGB_dCS_PRL_Nair2019}, one needs to divide the result of Ref.~\citep{QNM_arxiv_Carullo2021} by a factor of approximately $4$. 
We are grateful to the anonymous referee for pointing out this relation. }
The analysis based on inspiral data is different.
The leading order of the scalar charge effect in \ac{EdGB} gravity enters at $-1$ \ac{PN} while the leading order of the tidal effect enters at $1$ \ac{PN} \citep{tidal_PRD_Vines2011, tidal_PED_Damour2012}, which means that the modification on the inspiral waveform keeps the original form in this theory. 
The only difference is $s^{\rm EdGB}=0$ for \ac{NS} components. 
However, constraining the \ac{EdGB} under the assumption that GW190814 is a \ac{NSBH} is not feasible since there is no \ac{NSBH} waveform model that includes both higher multipoles and precession effects. 
If these two effects are excluded, we cannot unbiasedly constrain the secondary mass of GW190814 as well \citep{GW190814_APJL_LIGO2020}, and then the constraints on \ac{EdGB} and \ac{dCS} gravity would be biased. 
Note that in the discussion below Eq.(\ref{eq:beta_dCS}) we have presented some arguments for the \ac{BBH} merger origin of GW190814. 

\begin{table}
    \begin{center}
    \setlength{\tabcolsep}{1pt} 
    \renewcommand{\arraystretch}{1.2} 
        \begin{tabular}{@{\extracolsep{4pt}}
        c|c|c@{}}
            \hline
            \hline
            Events& Methods& $\sqrt{\alpha_{\rm EdGB}}\,(\rm km)$  \\
            \hline
            \multicolumn{1}{c|}{\multirow{2}{*}{GW190412}}&
            Reweighting & $4.41$ \\
            \multicolumn{1}{c|}{}&
            Bayes XPHM & $4.46$ \\
            \hline
            \multicolumn{1}{c|}{\multirow{3}{*}{GW190814}}&
            Reweighting & $0.93$ \\
            \multicolumn{1}{c|}{}&
            Bayes XPHM & $0.40$ \\
            \multicolumn{1}{c|}{}&
            Bayes Pv3HM & $0.38$ \\
            \hline
            \hline
        \end{tabular}
    \end{center}
\caption{
Constraints on $\sqrt{\alpha_{\rm EdGB}}$ with reweighting and Bayesian analyses. 
The Bayesian analyses are performed with both the IMRPhenomPv3HM (Pv3HM) waveform model and the IMRPhenomXPHM (XPHM) waveform model. 
All constraints are presented as the upper limits for each individual event at $90\%$ credibility. 
}
\label{table:compare}
\end{table}

\begin{figure*}
\centering
\begin{subfigure}[b]{0.5\linewidth}
\centering
\includegraphics[width=\textwidth,height=8cm]{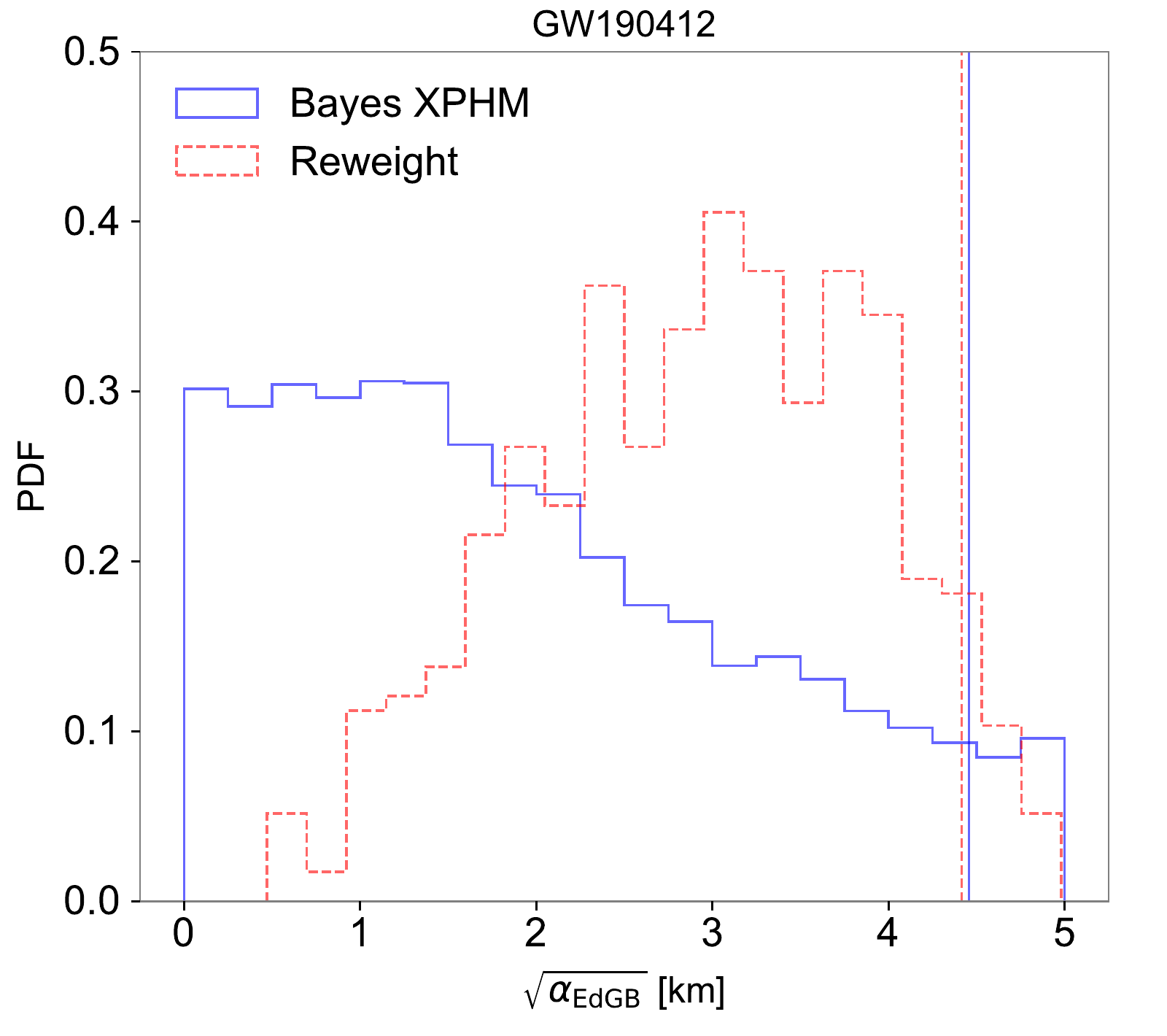}
\end{subfigure}%
\begin{subfigure}[b]{0.5\linewidth}
\centering
\includegraphics[width=\textwidth,height=8cm]{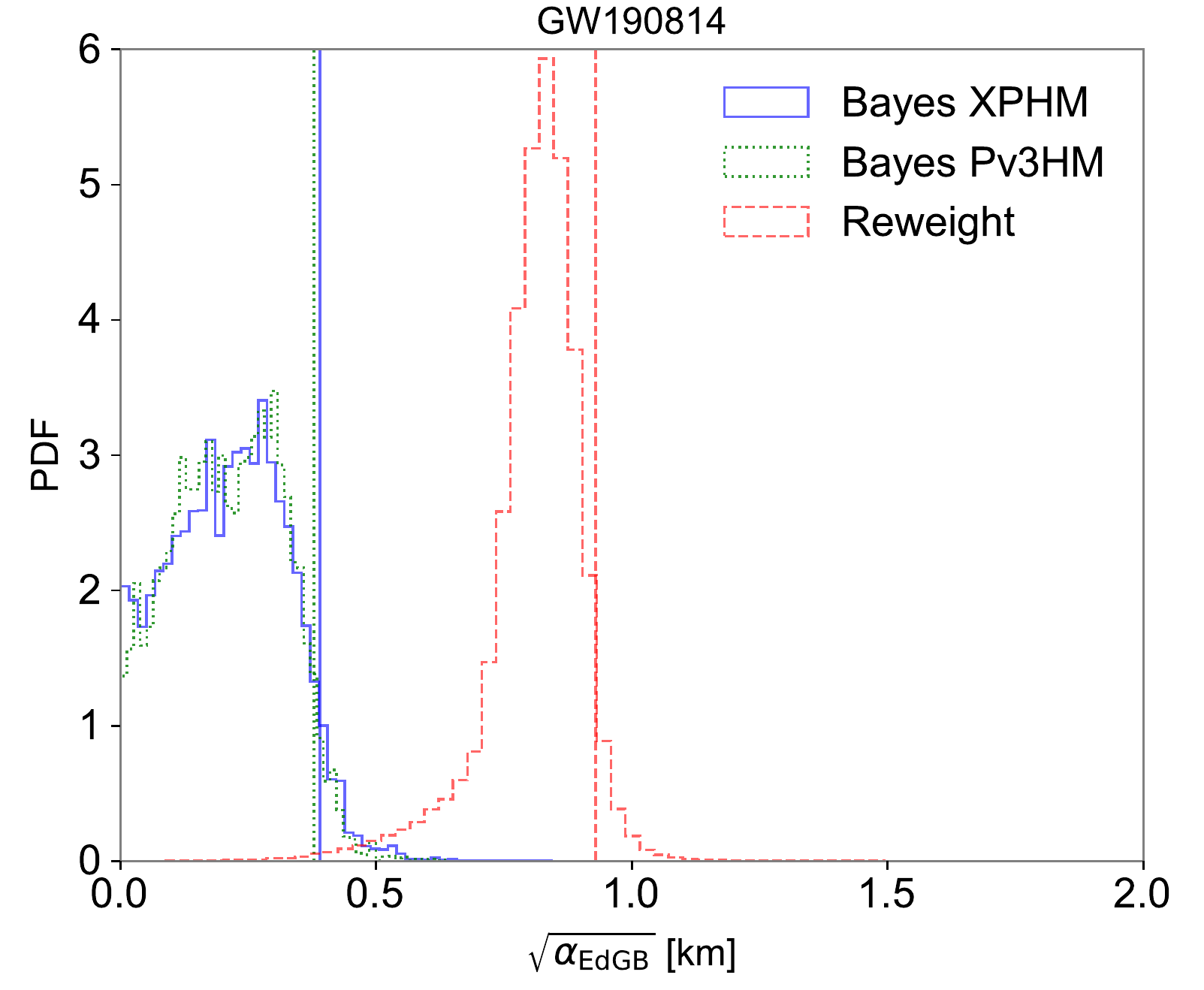}
\end{subfigure}%
\caption{
Probability density functions of $\sqrt{\alpha_{\rm EdGB}}$ for GW190412 (left panel) and GW190814 (right panel). 
For GW190412, the constraints of the Bayesian analysis performed with the IMRPhenomXPHM waveform model (solid histogram) and the reweighting analysis (dashed histogram) are $\sqrt{\alpha_{\rm EdGB}}\lesssim 4.46$ and $\sqrt{\alpha_{\rm EdGB}}\lesssim 4.41\,\rm km$, at $90\%$ credible level, respectively. 
For GW190814, the constraints from Bayesian inference with the IMRPhenomXPHM (solid histogram) waveform model is $\sqrt{\alpha_{\rm EdGB}}\lesssim 0.40\,\rm km$ at $90\%$ credibility. 
For GW190814, the constraints from Bayesian inference with both IMRPhenomXPHM (solid histogram) and IMRPhenomPv3HM (dotted histogram) waveform models are $\sqrt{\alpha_{\rm EdGB}}\lesssim 0.40$ and $\sqrt{\alpha_{\rm EdGB}}\lesssim 0.38\,\rm km$ at $90\%$ credible level, respectively. 
As a comparison, the reweighting analysis (dashed histogram) gives the constraint $\sqrt{\alpha_{\rm EdGB}}\lesssim 0.93\,\rm km$ at $90\%$ credibility. 
For both events, results of the reweighting analysis are biased, leading to evidence for deviation from \ac{GR}. 
This is due to the absence of the correlations between parameters in this method. 
}
\label{fig:EdGB_posteriors}
\end{figure*}

We further carry out Bayesian analyses on GW190412 and on GW190814, assuming they are \ac{BBH} events, which give better constraints on both $\sqrt{\alpha_{\rm dCS}}$ and $\sqrt{\alpha_{\rm EdGB}}$ parameters. 
Among them, the posteriors of $\sqrt{\alpha_{\rm dCS}}$ are still not respecting the small-coupling conditions, and thus cannot place meaningful constraint on it. 
The comparison of constraints between the fully Bayesian and the reweighting analyses on $\sqrt{\alpha_{\rm EdGB}}$ is summarized in Table~\ref{table:compare}. 
In Fig.~\ref{fig:EdGB_posteriors}, we also show the posterior distributions of $\sqrt{\alpha_{\rm EdGB}}$, which are from the Bayesian and reweighting analyses of GW190412 and GW190814. 

For GW190814, the Bayesian inferences are performed with both IMRPhenomPv3HM and IMRPhenomXPHM waveform models, resulting in similar constraints, $\sqrt{\alpha_{\rm EdGB}}\lesssim 0.40\,\rm km$ and $\sqrt{\alpha_{\rm EdGB}}\lesssim 0.38\,\rm km$ at $90\%$ credibility, respectively. 
The little difference between results of these two waveform models can be explained by the statistical noise. 
Thus, the difference between these two waveform models can be ignored in this work. 
On the other hand, the constraint from the reweighting analysis is $\sqrt{\alpha_{\rm EdGB}}\lesssim 0.93\,\rm km$ at $90\%$ credibility. 
The Bayesian-analyses-based constraints are tighter than that of the reweighting analysis by a factor of approximately $2$. 
Also, notice that more than $99\%$ of the posterior distributions of GW190412 and GW190814 from the reweighting analysis lies beyond $0$, which may mislead deviations from \ac{GR}. 
Differences between the reweighting analysis and the Bayesian analysis are reasonable since the reweighting analysis does not take into account full correlations between parameters as the Bayesian analysis does. 
Strictly speaking, we employ resample points from the posterior distributions of parameters $(\delta\phi_{-2}, m_1, m_2, \chi_1, \chi_2, \theta_1, \theta_2)$ individually in order to get the posterior of $\sqrt{\alpha_{\rm EdGB}}$, which may also lead to a biased posterior distribution on it. 

For GW190412, the constraint from the Bayesian analysis is similar to that from the reweighting analysis at $90\%$ credibility, but the former performs better than the latter at $68.3\%$ credibility, $\sqrt{\alpha_{\rm EdGB}}\lesssim 2.46$ and $\sqrt{\alpha_{\rm EdGB}}\lesssim 3.53\,\rm km$ respectively. 
The constraint from the Bayesian analysis on GW190412 is not as stringent as that of GW190814, which is due to the shorter inspiral signal (lower SNR) and the smaller mass ratio (assuming $q>1$) of GW190412. 
Thus, the difference between results of a Bayesian analysis and the reweighting analysis is not as obvious as GW190814. 
The posterior distributions for all parameters of GW190412 and GW190814 are presented in Appendix.~\ref{apd:posteriors}. 

\section{Conclusions}\label{sec:summary}
In this work, we have tried to constrain the parameters of two typical theories alternative to \ac{GR}, \ac{EdGB} and \ac{dCS}, $\sqrt{\alpha_{\rm EdGB, dCS}}$, by adopting two methods, i.e., the reweighting and full Bayesian analyses to analyze the two events of GWTC-2, GW190412, and GW190814. 
As expected, these two events in GWTC-2 cannot place constraints on the parameter $\sqrt{\alpha_{\rm dCS}}$. 
A $90\%$ bound of $\sqrt{\alpha_{\rm EdGB}}\lesssim 0.40\,\rm km$ is given by Bayesian analysis of GW190814, assuming it is a \ac{BBH} merger. 
The constraint from Bayesian analysis of GW190412 is $\sqrt{\alpha_{\rm EdGB}}\lesssim 4.46\,\rm km$ at $90\%$ credibility. 
The Bayesian analyses give a stringent constraint on $\sqrt{\alpha_{\rm EdGB}}$, in agreement with \ac{GR}, and this is different from the posterior distributions given by the reweighting analysis. 
The results of the reweighting analysis are biased, which may attribute to the fact that the reweighting analysis ignores the correlations between parameters. 

Before this work, the most stringent constraint ($90\%$ credibility) is $\sqrt{\alpha_{\rm EdGB}}\lesssim 0.8\,\rm km$ \citep{QNM_arxiv_Carullo2021}, which was obtained from the stability analysis of BHs in \ac{EdGB} gravity. 
The constraint from GW190814 has improved it by a factor of about $2$. 
Note that the constraint presented by \citet{QNM_arxiv_Carullo2021} does not rely on a perturbative assumption, while our result does. 
In Ref.~\citep{EdGB_dCS_PRL_Nair2019}, a $90\%$ bound of $\sqrt{\alpha_{\rm EdGB}}\lesssim 5.6\,\rm km$ was reported by analyzing the GW events in GWTC-1. 
The fact that the constraints from GW190814 have been improved by about an order of magnitude is reasonable due to the presence of precession and higher multipoles, which will improve the estimation of masses and spins of the sources \citep{GW190412_PRD_LIGO2020, GW190814_APJL_LIGO2020}. 
From the fundamental physics side, the strongest constraint we have placed on \ac{EdGB} gravity may strengthen the pillars of \ac{GR} such as the strong equivalence principle and the no-hair theorem of black holes.

\begin{acknowledgments}
H.T.W. is grateful to J.L.J. for insightful comments and discussions. 
This work has been supported by NSFC under Grants No. 11921003,  No. 11847241, No. 11947210, and No. 12047550. 
P.C.L. is also funded by China Postdoctoral Science Foundation Grant No. 2020M67001. 
H.T.W. is also supported by the Opening Foundation of TianQin Research Center. 
This research has made use of data, software, and/or web tools obtained from the Gravitational Wave Open Science Center (https://www.gw-openscience.org), a service of LIGO Laboratory, the LIGO Scientific Collaboration, and the Virgo Collaboration. 
LIGO is funded by the U.S. National Science Foundation. 
Virgo is funded by the French Centre National de Recherche Scientifique (CNRS), the Italian Istituto Nazionale della Fisica Nucleare (INFN), and the Dutch Nikhef, with contributions by Polish and Hungarian institutes.

\end{acknowledgments}

\appendix

\section{Posterior distributions}\label{apd:posteriors}
\begin{figure*}
\centering
\includegraphics[width=0.95\linewidth]{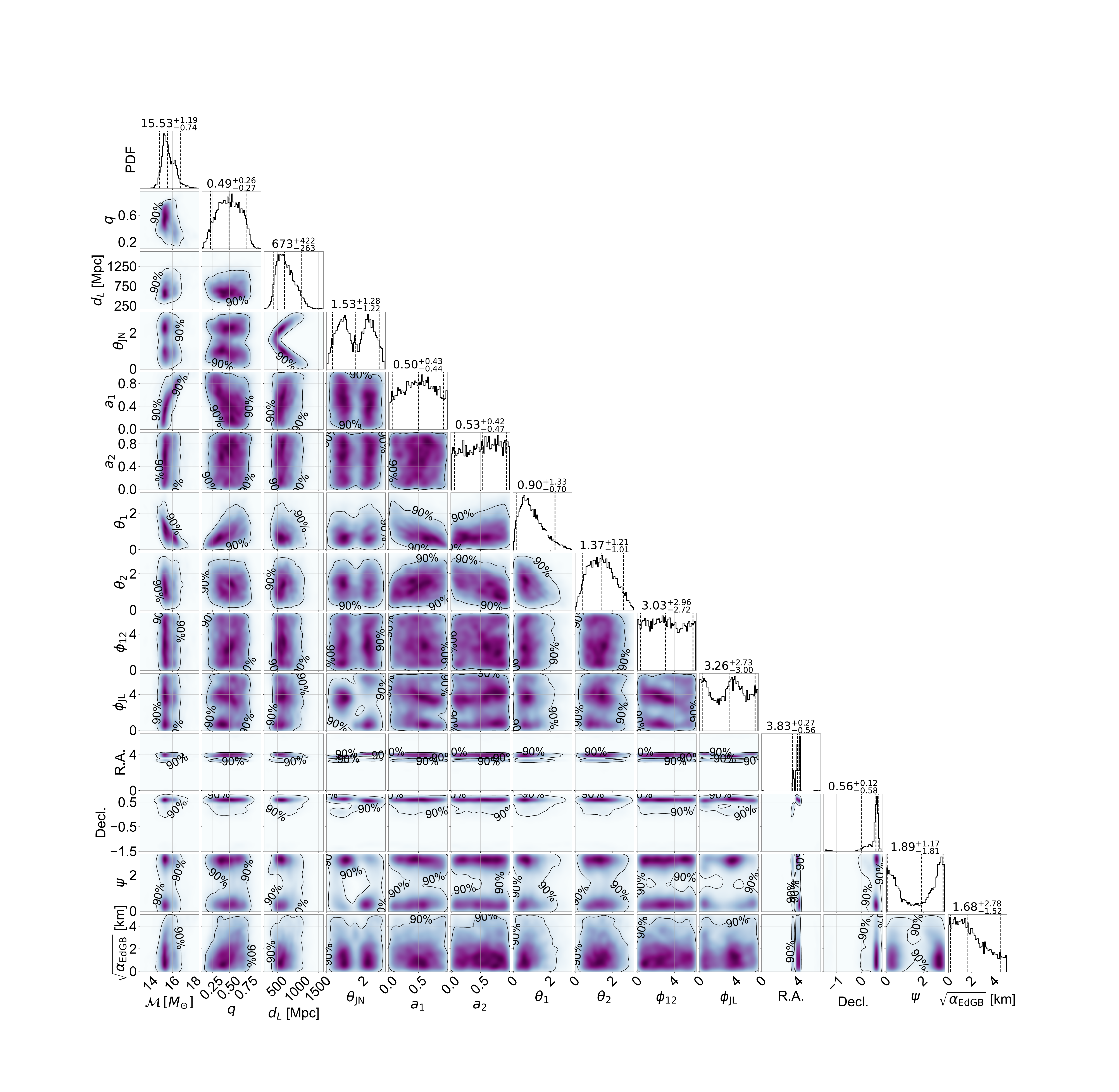}
\caption{
Posterior distributions for parameters of GW190412, analyzed with the IMRPhenomXPHM waveform model. 
The low-pass cutoff frequency $f_{\rm high}$ is $83$ Hz. 
The contours represent the credible level at $90\%$. 
The numbers listed in the diagonal are the median values and the $90\%$-credible measurements of each parameter. 
}
\label{fig:190412_full}
\end{figure*}

\begin{figure*}
\centering
\includegraphics[width=0.95\linewidth]{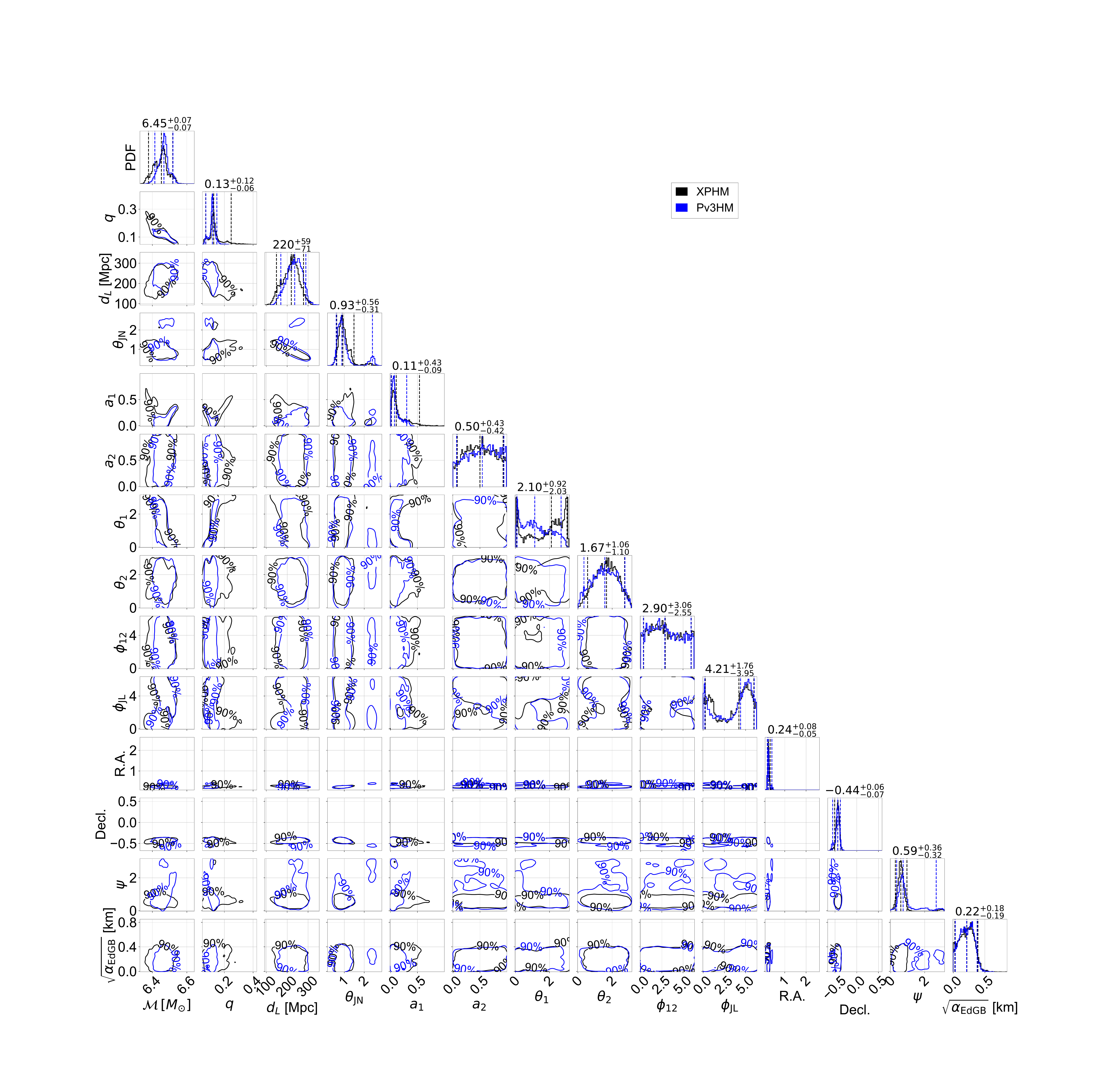}
\caption{
Posterior distributions for parameters of GW190814, analyzed with IMRPhenomXPHM and IMRPhenomPv3HM waveform models. 
The low-pass cutoff frequency $f_{\rm high}$ is $137$ Hz. 
The contours represent the credible level at $90\%$. 
The numbers listed in the diagonal are the median values and the $90\%$-credible measurements of each parameter of the IMRPhenomXPHM waveform model. 
}
\label{fig:190814_full}
\end{figure*}

In Figs.\ref{fig:190412_full} and \ref{fig:190814_full}, we show the posterior distributions for all parameters of GW190412 and GW190814, respectively. 
The ranges of these distributions are wider compared with that of full \ac{IMR} analyses \citep{LIGO_O3a_arxiv2020}, which is due to the low-pass cutoff frequencies $f_{\rm high}$ used in analyses of GW190412 ($f_{\rm high}=83$ Hz) and GW190814 ($f_{\rm high}=137$ Hz). 

\bibliographystyle{apsrev4-1}
\bibliography{EdGB_dCS}

\end{document}